\documentclass[acmlarge]{acmart}

\makeatletter
\newcommand{\confshort}{\acmConference@shortname}
\newcommand{\conffull}{\acmConference@name}
\newcommand{\confdate}{\acmConference@date}
\newcommand{\confloc}{\acmConference@venue}
\AtBeginDocument{
  \fancypagestyle{firstpagestyle}{
    \fancyhead{}%
    \fancyfoot[C]{}%
  }
  \fancyhf{}
  \fancyhead[LO]{\@headfootfont\shorttitle}%
  \fancyhead[RE]{\@headfootfont\@shortauthors}%
  \fancyhead[LE]{\@headfootfont\footnotesize \confshort, \confdate, \confloc}%
  \fancyhead[RO]{\@headfootfont\footnotesize \confshort, \confdate, \confloc}%
  \fancyfoot[C]{}%
}
\makeatother
\acmBooktitle{\conffull\@ (\confshort), \confdate, \confloc}

\setcopyright{acmlicensed}
\copyrightyear{2026}
\acmYear{2026}
\setcopyright{cc}
\setcctype{by}
\acmConference[FAccT '26]{The 2026 ACM Conference on Fairness, Accountability, and Transparency}{June 25--28, 2026}{Montreal, QC, Canada}
\acmBooktitle{The 2026 ACM Conference on Fairness, Accountability, and Transparency (FAccT '26), June 25--28, 2026, Montreal, QC, Canada}
\acmDOI{10.1145/3805689.3812307}
\acmISBN{979-8-4007-2596-8/2026/06}

\acmBooktitle{The 2026 ACM Conference on Fairness, Accountability, and Transparency (FAccT’26), June 27–30, 2026, Montréal, Canada}

\usepackage{tabularx}
\begin{document}

\title{The Fair Lending Model: How the Longest-Running Algorithmic Fairness Programs Work in Practice}

\author{Emily Black}
\authornote{Equal contribution}
\email{emilyblack@nyu.edu}
\affiliation{%
  \institution{New York University}
  \city{New York}
  \country{USA}
}

\author{Miranda Bogen}
\authornotemark[1]
\email{mbogen@cdt.org}
\affiliation{%
  \institution{Center for Democracy \& Technology}
  \city{Washington, DC}
  \country{USA}
}

\author{Logan Koepke}
\authornotemark[1]
\email{logan@upturn.org}
\affiliation{%
  \institution{Upturn}
   \city{Washington, DC}
  \country{USA}
}

\author{Solon Barocas}
\email{solon.barocas@microsoft.com}
\affiliation{%
  \institution{Microsoft Research}
  \city{New York}
  \country{USA}
}

\author{Wesley Deng}
\email{hanwend@andrew.cmu.edu}
\affiliation{%
  \institution{Carnegie Mellon University}
  \city{Pittsburgh}
  \country{USA}
}

\author{Mingwei Hsu}
\email{ming@upturn.org}
\affiliation{%
  \institution{Upturn}
   \city{Washington, DC}
  \country{USA}
  }

\renewcommand{\shortauthors}{Black, Bogen, Koepke, Barocas, Deng, and Hsu}
\renewcommand{\shorttitle}{The Fair Lending Model}

\begin{abstract}
U.S. financial institutions subject to fair lending laws have been running algorithmic fairness programs for decades. Despite this long history, remarkably little is known about how these requirements operate in practice. In this paper, we offer the first empirical account of how financial institutions test for and mitigate algorithmic discrimination on the ground. In doing so, we shed light on how the regulatory design of fair lending law and regulation have shaped the policies, processes, and practices of fair lending programs. Drawing on 35 semi-structured interviews with participants across the fair lending ecosystem, we find that while financial institutions have a floor of fairness practices aimed at preventing discrimination in lending largely absent in other domains, the specifics of how firms test for discrimination and search for less discriminatory algorithms varies widely. 
We also find that regulatory supervision via fair lending examinations has been the key driver of compliance work, but that the practical impact of fair lending programs often depends on how well they can navigate competing business incentives, perceived legal tensions, and regulatory uncertainty. Ultimately, our findings highlight the unique role that supervisory authority has played in successfully fostering fair lending practices---a regulatory design feature that is distinct from other areas of civil rights law and almost completely absent from recent policy proposals for dealing with algorithmic discrimination.
\looseness=-1

\end{abstract}



\authorsaddresses{Corresponding author: Emily Black, emilyblack@nyu.edu, NYU, New York, USA}
\maketitle

\section{Introduction}
\label{sec:intro}
For over a decade, researchers have worked to draw attention to the harms resulting from algorithmic systems and have developed mechanisms to measure and mitigate a variety of these harms, including discrimination~\cite{dwork2012fairness,chouldechova2017fair,hardt2016equality,agarwal2018reductions, caton2024fairness}. 
More recently, a line of empirical scholarship has begun to examine the challenges faced by industry practitioners working on responsible AI in practice to do so, particularly at large technology companies~\cite{madaio2020co,deng2022exploring,deng2023understanding,deng2023investigating}. 

Meanwhile, for nearly 50 years, U.S. 
financial institutions have been under a legal obligation to not discriminate in credit transactions~\cite{ecoa1974,fha}. For the vast majority of that time, those same institutions have relied upon algorithmic decision-making systems to determine who gets access to credit and on what terms~\cite{frb2007creditscore,capon1976creditscoring,poon2012whatlenderssee}.
As a result, fair lending programs---programs to ensure that financial institutions comply with fair lending laws---are likely the longest-running example of algorithmic fairness on the ground. This overlooked history offers crucial lessons for the FAccT community, from both a technical and policy standpoint. Our paper is the first to empirically document 
the work of fair lending programs, and does so at a time of unprecedented erosion of fundamental building blocks of fair lending protection.

A range of scholars have offered a robust picture of potential methods to test for and mitigate discrimination and detailed the 
struggles practitioners face
when attempting to do so \cite{balayn2023fairness,zhang2024ai,ryan2023integrating, deng2023investigating, madaio2024tinker, smith2025pragmatic}, but few of these studies have looked specifically at the role that formal regulations play in structuring those practices. The long history of fair lending programs offers a critical window into the way regulatory design choices---that is, how a regulatory system is structured \cite{baldwin2011understanding}---can shape algorithmic fairness efforts in practice. 
Understanding how regulatory design can materially influence industry efforts to test for and mitigate discrimination is critically important, especially given the increasing focus on legal and regulatory mechanisms, tools, and instruments that may be necessary to advance algorithmic fairness.

Our study of fair lending programs at U.S. financial institutions focuses on three main questions:

\textbf{RQ 1:} How do fair lending teams at U.S. financial institutions test for and mitigate discrimination in algorithmic systems?

\textbf{RQ 2:} How does the regulatory design of fair lending law influence the policies, processes, and practices we observe in fair lending teams in U.S. financial institutions?

\textbf{RQ 3:} 
What challenges do fair lending practitioners face in ensuring that discrimination is 
effectively reduced?

To answer our research questions, we conducted 35 semi-structured interviews with participants across the U.S. financial industry, including engineers, lawyers, regulators, and third-party vendors. \textbf{Ultimately, this paper offers the first empirical account of how financial institutions test for and mitigate algorithmic discrimination on the ground. 
In doing so, we also shed light on how the regulatory design of fair lending law and regulation has shaped the policies, processes, and practices of fair lending teams.}

For efforts to advance algorithmic fairness to be successful at both a methodological and structural level, we argue they must be rooted in evidence on how regulatory design impacts algorithmic fairness work on-the-ground. Such evidence is of particular importance now, as policymakers across the country are introducing both legislative and regulatory proposals with the aim of better combating algorithmic discrimination 
~\cite{colorado_ai_act2024,texas_responsible_ai_act2025,california_sb53_2025,nyc_local_law_144_2021,california_ab1018_2025,newyork_s1962_2025,us_s3308_2025}. Ultimately, we believe that by understanding what firms have done---and specifically by understanding how those efforts were informed and shaped by specific regulatory design choices---advocates, practitioners, policymakers, and scholars will be better positioned to understand what regulatory conditions are needed to reduce algorithmic discrimination and why. While we set out to elicit insights that could support technical and policy efforts to reduce  discrimination in other consequential contexts, our findings ultimately also serve to document the financial industry's practices immediately leading up to the current U.S. administration's efforts  to actively dismantle the very regulatory regime of interest.

\section{Background}
\label{sec:background}
Four laws collectively serve as the foundation for fair lending regulation in the U.S.: The Equal Credit Opportunity Act (ECOA), the Fair Housing Act (FHA), the Home Mortgage Disclosure Act (HMDA) and the Community Reinvestment Act (CRA) \cite{ecoa1974, fha, hmda1975, cra1977}. Passed in 1968, the FHA prohibits discrimination when renting or buying a home, getting a mortgage, or seeking housing assistance \cite{fha}. ECOA (1974) forbids discrimination in any aspect of a credit transaction on the basis of race, sex, age, and other protected characteristics \cite{ecoa1974}. Both the FHA and ECOA prohibit  disparate treatment, which emphasizes discriminatory intent, and practices that have a disparate impact.\footnote{Since data collection, the CFPB finalized a rule removing disparate impact liability from Regulation B, the regulation implementing ECOA. This changes the agency's regulatory posture but does not resolve whether ECOA itself prohibits disparate impact, which courts have consistently found it does.}
HMDA (1975) requires financial institutions to maintain, report, and publicly disclose loan-level information about mortgages, in order to help identify potential discriminatory lending patterns \cite{hmda1975}. The CRA (1977) is designed to ensure that mortgage lenders serve low- and moderate-income neighborhoods, thereby combating practices like redlining \cite{cra1977}. Notably, only ECOA and the FHA impose substantive legal obligations on financial institutions to not discriminate. HMDA, meanwhile, compels the production of data that plaintiffs and regulators can rely upon to identify discriminatory patterns, whereas the CRA imposes a general obligation for lenders to meet the needs of their community. 

The regulatory design of fair lending law is strikingly different from other civil rights laws. By regulatory design, we mean the ways in which a regulatory system is structured---from the institutional form of the regulator to its enforcement tools to its compliance mechanisms---to achieve its stated goals \cite{baldwin2011understanding, cohen2025policymaking}. To understand why, consider the main ways in which Title VII, the foundational federal law prohibiting employment discrimination, is enforced. First, aggrieved individuals who believe they have been victims of discrimination must file a complaint with the Equal Employment Opportunity Commission (EEOC) \cite{cfr29part1691}. The EEOC will investigate these complaints and either file their own lawsuit or issue a Notice of Right to Sue, allowing the aggrieved individual to vindicate their private right of action and file suit against an employer for discrimination \cite{cfr29part1691}. The EEOC is almost completely reactive: to enforce Title VII, it almost exclusively relies upon aggrieved individuals filing a complaint of discrimination \cite{walter1995fairlending}. 

Fair lending laws like the FHA and ECOA also equip victims of discrimination with a private right of action to sue lenders for discriminatory actions and empower regulators to investigate complaints. However, financial regulators who enforce fair lending laws---like the Consumer Financial Protection Bureau (CFPB)---also possess robust statutory authorities that authorize them to proactively and routinely inspect business practices for compliance with fair lending laws, \textit{whether or not the regulator believes discriminatory practices are occurring} \cite{walter1995fairlending}. Such supervisory authority and examination, where regulators actively oversee regulated entities’ compliance by sending examiners to a financial institution’s offices to probe for evidence of lending discrimination \cite{cohen2024regulatory}, are the key to the regulatory design of fair lending law. 
Through supervision and examination, fair lending regulators can direct financial institutions to address potential deficiencies in antidiscrimination efforts. If such deficiencies are not resolved, they can be elevated to formal enforcement proceedings. 
Ultimately, “supervision is where the practice of government actually occurs in finance” \cite{contibrown2025}.

Finally, this regulatory regime only applies to legally defined ``covered entities.'' These include institutions that directly furnish credit to consumers, but does not include, for instance, vendors of credit scores \cite{selbst2023unfair}. In the remainder of the paper, we refer to entities as either ``covered entities'' or ``entities not covered'' to indicate whether they are subject to these laws. We note that some elements of these laws, regulatory regimes, and supervisory priorities are in active flux in response to executive actions under the current U.S. administration.

\section{Related Work}
\label{sec:related}
A growing body of research in fields such as Human-Computer Interaction (HCI) and Responsible AI (RAI) has investigated how industry practitioners engage in fairness work within real-world organizational contexts \cite{holstein2019improving, orr2020attributions, ali2023walking, widder2024power, deng2023understanding, passi2019problem, madaio2020co, madaio2024tinker, smith2023many, wang2023designing, rakova2021responsible, deng2025supporting}.
Several works have revealed how practitioners evaluate model fairness in their organizations, 
identifying technical challenges in bias mitigation in practice, like selecting appropriate metrics and determining which stakeholder groups to prioritize~\cite{madaio2020co}, and studying use and misuse of popular open-source fairness tools~\cite{deng2022exploring, balayn2023fairness, smith2023many}. 
More generally, many have emphasized the difficulties in concretizing abstract responsible AI best practices or principles, suggesting these soft requirements may not often lead to more responsible systems in practice~\cite{akbar2024trustworthy, bughin2024doing, gunasekara2025systematic, ali2023walking, madaio2024learning, sadek2024challenges, smith2025pragmatic, widder2023dislocated, madaio2024tinker}. While a few works have touched on the impacts of regulatory requirements around protecting user privacy~\cite{lee2024don, sarathy2023don}, none have examined fairness practices in organizations explicitly covered by civil rights law, which has required the mitigation of unlawful disparities in credit, housing, and employment contexts for decades. Similarly, while others have provided a law and policy overview of fair lending considerations for machine learning models and algorithmic fairness \cite{kumarecoa,hall2021fair}, these works do not detail what financial institutions do in practice. More recent work outside the academy has surveyed relevant fair lending techniques \cite{finreglab2023,finreglab2023ml}, but has not considered how regulatory design has shaped compliance practices.

An emerging body of work within the FAccT community has sought to evaluate the efficacy of new regulations targeting algorithmic discrimination in employment, specifically New York City’s Local Law 144 \cite{10.1145/3630106.3658998,10.1145/3630106.3658959,lam2024framework,10.1145/3715275.3732004}. These studies have largely focused on establishing the degree to which covered actors appear to be complying with the law, assessing compliance quality, and identifying weaknesses in the law that might account for its lack of efficacy. Unlike our study, these works focus on newly implemented regulations, rather than well-established civil rights laws.

Additionally, little prior work has directly engaged compliance practitioners as study participants, in part because many existing studies have not focused on highly regulated domains like the financial sector, where compliance requirements can strongly shape  fairness testing and mitigation practices \cite{nichol2024moral, cinca2025practitioners, akbarighatar2024operationalizing}. Many prior studies have acknowledged as a limitation the challenge of recruiting compliance professionals, who, given the sensitivity of their work, are often harder to access than technical practitioners \cite{deng2023investigating, madaio2024tinker, wang2022whose, sambasivan2022deskilling}. Our work extends the literature by being among the first to investigate how fairness testing and mitigation are carried out in practice in the highly regulated domain of lending, extending empirical insights beyond engineers or researchers to include lawyers, third-party vendors, and regulators.

Furthermore, while many empirical studies of RAI in practice highlight regulation as a key lever for motivating and scaffolding fairness work \cite{deng2023understanding, widder2022limits, orr2020attributions, rakova2021responsible, ryan2024ai}, the FAccT community still lacks an in-depth empirical understanding of how regulatory design can shape firms' efforts and  practitioners’ work. This gap is especially consequential for fairness testing, where existing policies and regulatory guidance are frequently described as incomplete, difficult to operationalize, or misaligned with practitioners’ day-to-day workflows \cite{madaio2024tinker, madaio2024learning, deng2023investigating, ali2023walking}. Through an in-depth interview study with practitioners across roles, our work offers concrete insights about how details of regulatory design can significantly shape the antidiscrimination testing done---and not done---inside corporations. 
Ultimately, our work is perhaps most in line with prior sociolegal studies of firms seeking to comply with discrimination and privacy law \cite{edelman2020working,bamberger2015privacy,waldman2021industry}, which find that firms are able to take advantage of significant latitude in how to interpret regulation---and often do so in ways that substitute performative compliance practices for meaningful changes in corporate behavior.

\section{Methods}
\label{sec:methods}

\noindent\textit{\textbf{Study Design.}} We conducted semi-structured interviews with 35 participants related to financial services in the U.S. We developed different interview protocols for our four main populations: engineers, lawyers, third parties, and regulators. We first conducted a pilot study with six participants to help construct effective protocols for each role. Once protocols were in place (see Appendix  \ref{app:protocol_policy}, \ref{app:protocol_engineer}, and \ref{app:protocol_regulator}), we conducted semi-structured interviews to understand participants’ current practices and challenges around testing and mitigating discrimination in algorithmic systems. Semi-structured interviews each lasted up to an hour. The interview began by asking participants to describe their current role, their previous relevant experiences, and their organization's process and structure as it relates to algorithmic fairness. Then, we tailored  questions depending on the population. For example, we asked engineers about techniques and metrics, while we asked lawyers how disparate impact doctrine informs fairness testing in their organization. 

\noindent\textit{\textbf{Participants.}} We adopted a purposive sampling approach with the goal of recruiting practitioners across multiple roles who had experience testing and mitigating discrimination in algorithmic systems broadly deployed in the financial services sector. We reached out to more than 100 practitioners through a combination of direct outreach and snowball sampling, leveraging the research team’s professional networks and referrals. Ultimately, 35 participants, spanning dozens of organizations, took part in the study. Table \ref{tab:participants} in the Appendix provides an overview of these participants.  
We note that we refer to participants as either E, L, T, or R depending on whether their role matched most closely with an engineer, lawyer, third-party actor, or regulator.

Given the sensitivity of fair lending and compliance-related work, we took several steps to protect participant confidentiality. Following prior work on responsible AI practices in industry, we omitted demographic details and abstracted certain information about participants’ employers and roles to avoid inadvertently identifying individuals working in this high-stakes domain. In addition, we assured participants that we would not ask them to reveal any confidential or personally identifying information about themselves or their colleagues and that we would de-identify all responses at the individual, team, and organization levels. 
All participants were offered a \$50 honorarium for their time, though many declined due to organizational policies. Our protocol was approved by our institutions' IRB. \looseness=-1

\noindent\textit{\textbf{Data Analysis.}} Our study sessions yielded approximately 38 hours of audio that we transcribed. To analyze our interview transcripts, we adopted a reflexive thematic analysis approach \cite{braun2019reflecting}. Members of our research team met after each interview session to conduct an interpretation session and collaboratively develop a codebook for annotation, which we include in the Appendix. We then distributed the transcripts among the research team for open coding. 
Each transcript was analyzed by one team member, typically the person who led the interview. Another team member then reviewed and validated the interpretations made by the primary annotator. 
Throughout the coding process, the authors regularly discussed discrepancies in interpretation and iteratively refined the codes based on these discussions. Consistent with reflexive thematic analysis, we did not calculate intercoder reliability, as coding is understood to be interpretive and iterative rather than aimed at statistical agreement
 \cite{braun2019reflecting}. \looseness=-1  In total, we generated around 1,200 unique interpretation codes. Through an iterative, bottom-up affinity diagramming process, we grouped our interpretations into successively higher-level themes. We present our results in the following section, organized around our three research questions.

\section{Findings}
\label{sec:findings}
\subsection{RQ1: The contours of fair lending practice}

\label{sec:find:floor}
In this section, we describe the overall structure of discrimination testing and mitigation procedures by regulated financial institutions as explained by our participants. While participants broadly agreed that firms are expected to test for and prevent discrimination, and our data suggest that fair lending teams and formal antidiscrimination procedures are commonplace, the specific practices employed vary widely.

\subsubsection{Existence of Fair Lending Teams and Practices}
\label{sec:find:floor:exist}

As a first order matter, our interviews showed that fair lending teams invariably \emph{exist} within regulated financial institutions---which, as prior work has noted, is not always the case at firms that are not covered by fair lending law~\cite{madaio2020co}.
All of the individuals we interviewed who were at institutions covered by fair lending law described a designated fair lending unit that was responsible for monitoring compliance with relevant laws ($L_{5}$, $E_{1}$, $E_{6}$, $E_{4}$, $L_{4}$, $E_{11}$, $L_{2}$).  As $L_{5}$, who worked in such a department, described, “\textit{it is my responsibility to make sure that our bank does not discriminate against any of our customers or potential customers}.”

Beyond the existence of teams dedicated to antidiscrimination work, our interviews revealed a ubiquity and formality of fair lending procedures. 
Participants at regulated financial institutions almost universally believed they were expected to test models for potential discrimination, and as a result, had procedures to do so 
(e.g.\ $L_{2}$, $T_{1}$, $L_{3}$, $E_{4}$, $T_{5}$, $E_{6}$, $R_{1}$, $L_{5}$, $E_{8}$, $E_{9}$). 
As $L_{6}$
stated, “\textit{you need to test}" lending models for discrimination. Participants connected these procedures to regulatory requirements:  
$L_{2}$ said “\textit{The [Consumer Financial Protection] Bureau... [has] been clear that it’s a responsibility of the lender to make sure that their models are not causing disparate impact}."\footnote{In fair lending law, companies must prevent both disparate treatment (they may not make lending decisions on the basis of protected characteristics such as race) and disparate impact (they may not have underwriting procedures that result in disparities unless they can be justified by business necessity) ~\cite{ecoa1974}.}

While specific testing procedures varied between firms, a required procedure always existed, with many common elements. Many participants at regulated financial entities stated that a model would first be triaged to determine what level of bias testing would be pursued. For example, $L_{3}$ shared that ``\textit{we have a  policy that essentially allows for a model to be risk rated, depending on how it's being used and what it does, and then a testing process that is basically aligned with that risk rating}.'' The risk of a model is often related to the probability that its outputs would impact a consumer: $T_{9}$ shared, “\textit{the closer the model is to the consumer, the higher the fair lending risk}.”  $E_{9}$ noted that firms would conduct “\textit{a much deeper dive}” on models with the highest fair lending risk versus ``\textit{a little bit less in-depth analysis}” on other models. 

\subsubsection{Procedures to Test for Disparate Treatment and Disparate Impact} In order to prevent disparate treatment, participants  
($L_{2}$, $T_{1}$, $L_{3}$, $E_{4}$, $T_{5}$, $E_{6}$, $R_{1}$, $L_{5}$, $E_{8}$, $E_{9}$)
detailed multiple procedures to review individual model features---that is, tests to ensure that none of the input features to a model were themselves or served as an obvious proxy for a protected characteristic such as race. For example, $T_{1}$ described how most firms will “\textit{start with a qualitative variable review}” involving a review of a model’s features not only for protected characteristics (meaning, prohibited bases protected under state and federal law), but also whether the model’s features were “\textit{intuitively related to credit worthiness.}" Quantitative feature review would then involve assessing whether other features, even if not explicit prohibited bases themselves, were proxies that are “\textit{highly correlated with a prohibited basis}” ($L_{3}$). As part of both quantitative and qualitative feature review processes, $R_{1}$ observed that institutions often develop ``\textit{green and yellow and red lists for variables}'' where red features are ``\textit{not allowed in models},'' yellow warrant further analysis, and green have a clear and defensible relationship to credit.

Models that firms identify as posing the most fair lending risk---such as underwriting or pricing models---would then also be subject to a disparate impact review 
($L_{2}$, $T_{1}$, $T_{5}$, $T_{8}$, $L_{7}$, $T_{9}$, $T_{2}$): as $L_{7}$
shared, ``\textit{our highest [model] risk category [...] gets the deepest level of testing: a full disparate impact analysis.}'' As we discuss in more detail in the next section, we observed variation in how institutions defined disparate impact, as well as the thresholds at which a disparate impact triggered further analysis. Participants 
($T_{1}$, $T_{5}$, $T_{8}$)
explained that disparate impact testing relied on 
``\textit{tests like adverse impact ratios [(AIR)], standardized mean differences [(SMD)], [and] marginal effects}'' ($T_{8}$), with a focus on determining whether disparities were statistically and practically significant. Other participants examined disparities not at the model level in abstraction, but instead through the lens of historic outcomes. $L_{7}$’s institution would look at the default rates for various protected classes (e.g., the historic default rate for black and white borrowers) and then compare a model’s predictions against those historic outcomes to see if the model would worsen those disparities. If the model did not appear to worsen outcomes relative to historic rates, $L_{7}$ described that “\textit{we consider the model to have low fair lending risk}.” In other words, if the predicted default rate for specific subpopulations was in line with their historical base rate of default, the institution believed that it had a satisfactory business justification for the disparate impact.

Participants believed this variance was in part due to a lack of clear regulatory expectations on how such tests should be performed. $L_2$ noted there are “\textit{still no metrics or no concrete, crispy,  guidance}” from regulators. $E_{9}$ concurred, stating "\textit{there is no regulatory guidance that this is the exact way how you should be doing this disparity testing. So as long as it's reasonable [...] there is no black and white, right or wrong, answer here}." Conversely, clear regulatory expectations regarding methods to test for discrimination resulted in some standardization across industry, and a floor of practice. For example, the CFPB's whitepaper ~\cite{cfpb_bisg_whitepaper2014} indicating that Bayesian Improved Surname Geocoding (BISG) could be used to impute race information needed to perform discrimination testing offered many companies a tacit path to conducting discrimination testing despite institutions' trepidation around the collection of actual demographic data from customers.\footnote{
Some of this trepidation may stem from the law itself, which explicitly \textit{requires} discrimination testing when demographic data are collected~\cite{ecoa1974}.} 
 Many participants 
($T_{8}$, $R_{1}$, $R_{2}$, $T_{5}$, $T_{3}$, $L_{3}$, $T_{1}$, $E_{6}$)
described industry use of BISG as a standard.  As $T_{1}$ summarized, “\textit{I think that the CFPB issuing that BISG paper went a long way}" to making a free, easy, and public method supporting disparity measurement widely available, explaining that "\textit{it's not perfect, but we can all coalesce around} [it].” This contrasts with many studies of RAI in industry, which observe how constrained demographic data access can create friction for discrimination testing \cite{lee2024don, deng2023investigating, sarathy2023don, holstein2019improving}.

\subsubsection{Searching for Less Discriminatory Algorithms.} Following testing, many participants stated that if a model was found to exhibit disparate impact according to their firms' metrics, they would search for a less discriminatory algorithm (LDA)---that is, a model achieving a similar performance with less disparity 
($L_{2}$, $L_{3}$, $E_{1}$, $E_{4}$, $E_{6}$, $E_{11}$, $E_{9}$). As we discuss in  Section ~\ref{sec:find:no_consistent:tensions}, LDA search methods, and thus their probability of success, varied significantly across companies. 
Most participants reported \textit{drop-one} analysis~\cite{finreglab2023}---removing one input feature at a time and retraining---as their primary bias mitigation method.
However, a few participants (e.g., $E_1$) described more robust search processes that involved developing thousands of candidate models with different design choices and selecting the fairest among them within a 1-2\% performance tolerance ($T_{5}$, $E_{11}$).

\subsection{RQ2: Driving forces of discrimination work in regulated financial institutions}
\label{find:floor:risk}
We also engaged participants on the motivations that drove their antidiscrimination work, in order to understand how the \emph{regulatory design} of fair lending law structured their practice. In particular, we aimed to identify factors in the implementation and enforcement of fair lending law that seemed to specifically motivate institutions' policies, processes, and practices, as well as how they made compliance-related decisions. 
As a first order matter, our interviews revealed that a floor of practices exist in heavily regulated institutions that do not seem to reliably exist in  entities not covered, suggesting that the mere existence of fair lending regulation is a driving force for antidiscrimination work. As $L_{7}$ summed up, “\textit{if it wasn't for the fair lending legal risk, we wouldn't even do this review.}” More specifically, our participants shared that regulatory expectations and scrutiny were the main motivators behind the development and execution of fair lending work, as opposed to individual lawsuits brought through private right of action. 

\subsubsection{Participants cite regulatory expectations as a reason why their work was done at all.}
In our interviews, participants described how they would use regulatory expectations and fear of enforcement to internally motivate their work ($L_2$, $T_{5}$): as $T_{5}$ shared, for example, “\textit{Unfortunately, no one does this stuff out of the good and kindness of their heart. They do it because they're afraid of regulation}." $L_2$ expanded, noting how they would use enforcement actions to further motivate internal work, stating that “\textit{All the time, the fair lending team at [company name] would say, hey, this settlement just came out; you know we're doing pretty well in this, but we need to raise the bar because the settlement came out}.”
Conversely, participants shared that without regulatory expectations, antidiscrimination work would simply not happen ($R_{4}$, $L_1$). As $T_{10}$ stated about the domain of insurance, where they did not observe regular bias testing or mitigation, ``\textit{They have to be required to do it in order for them to do it [...] they're not gonna do it otherwise, in my opinion}.” As a further disincentive, $L_1$ explained that testing without a legal mandate could actually introduce additional legal risk to a firm.\looseness=-1

\subsubsection{Supervisory expectations and fair lending examinations as a primary motivator}
\label{sec:find:floor:authority}
While the general existence of fair lending law motivated firms' antidiscrimination efforts, participants were clear that supervisory expectations and fair lending examinations by regulators were the most consequential factors driving specific fair lending work at their firms, not private litigation.
For example, $L_{7}$ directly stated that “\textit{we spend more of our time worrying about [the] regulators, because we go through fair lending exams all the time}.” 

Regulatory scrutiny not only drove whether work was done, but also what specific work was done: $L_{7}$ suggested that “\textit{the regulators expect us to be searching for less discriminatory alternatives and documenting our searches},'' continuing to explain that their search process was ``\textit{well documented so that we can sort of show we've done our homework.}” $L_{7}$ ultimately concluded that, “\textit{if I was thinking about the motivations for why we take that action, it's really probably more to avoid an unpleasant fair lending exam than it is to worry too much about the litigation risk.}”

Regulatory expectations also shape what work was \emph{not} done. For example, $E_{6}$ noted that fair lending testing of credit models mostly meant testing for disparate impact on a limited set of protected characteristics (race and sex), despite the law covering multiple other protected classes: firms were “\textit{not very interested}'' in figuring out how to do tests for other protected characteristics “\textit{because from their perspective [...] the regulators are fine with what we’re doing}.”
\vspace{-0.5em}
\subsection{RQ3: Challenges to fair lending work in practice 
}
\label{sec:find:no_consistent}
Despite well-established fairness testing and mitigation processes, participants still described many challenges that complicated their ability to reduce disparate impacts. Here, we highlight three main challenges participants surfaced: first, the reality that the decision to follow the advice of the fair lending team is a business decision; second, that perceived legal tensions led to technical difficulties in mitigating disparities; and third, that uncertainty about regulatory requirements could be leveraged to justify the bare minimum to prevent discrimination. At times, though, this uncertainty seemed to provide some fair lending teams the 
flexibility to adopt surprisingly progressive policies and practices.

\subsubsection{Frustration with compliance being a business decision}
\label{sec:find:no_consistent:business}
Participants shared that the first challenge in avoiding the use of biased models in practice was convincing the business arm of the organization to do so.
They explained that deciding whether to 
take on the risk of deploying a model resulting in a disparity is in the hands of the business arm of an organization, which weighs the cost of compliance against the cost that would be incurred by a potential violation and the upside of proceeding, and decides whether to reduce the disparity. As $E_{6}$ described: "\textit{fair lending is a legal advisor}'' but the business ``\textit{can choose to take the advice and ignore it} [...] \textit{it is definitely within policy for an executive} [to] \textit{decide that }[an LDA] \textit{is not worth the cost, and they're gonna go ahead with }[the original model], \textit{but then they bet the entire legal risk of what it was to go ahead with.}”
Several participants described this dynamic as well
($L_{2}$, $L_{3}$, $T_{8}$, $E_{11}$, $L_{5}$, $E_{9}$).
As $L_{5}$ stated, there is little to no obligation to sacrifice some business objective in order to reduce disparity: ``\textit{I don't think there's an expectation that we have to make [...] financial decisions that are money losers. I don't think we have to make unprofitable decisions}.’’  

Beyond deciding whether or not to deploy an LDA, participants shared that the business arm of their firms were also in charge of ultimately deciding what counted as a viable LDA---in other words, how much leeway in model performance to grant in order to facilitate disparity reductions. The process of defining performance or cost thresholds for LDA viability was nearly always ad hoc, so decisions and practices varied widely: as $E_{4}$ observed, employees charged with model risk management (largely focused on "safety and soundness," or model performance, who might prefer minimal model degradation) and employees charged with fair lending (focused on reducing discrimination, and open to some minor model degradation if the alternate model resulted in meaningful narrowing of outcome gaps) would have to “\textit{hash it out}.” These conversations, and the need to constantly advocate for less discriminatory options, were a reported source of frustration for participants in fair lending teams: for example, $T_{5}$ shared that this question was “\textit{the source of half of my misery in life}.” $T_{8}$ shared that ``\textit{there's a lot of conflict}'' in these conversations.

Many participants reported that their firms had incredibly low tolerance for bearing performance costs to their models in order to reduce disparity: $L_2$ noted that  “\textit{modeling teams would be like no, the} [LDA model] \textit{has to perform equally as well or we’re not going to make the change},” where “equally as well” was interpreted very strictly. Similarly, $T_{9}$ noted that some modelers, when looking at a baseline model with a KS\footnote{The Kolmogorov-Smirnov (KS) metric measures the maximum separation of the cumulative distribution functions of the positive and negative classes of a model's predictions, i.e. defaulters and non-defaulters. This metric demonstrates the model's ability to meaningfully distinguish between classes.} of .5, and alternatives with a KS of .4, .48, .49, would suggest that “\textit{unless I get .5, it’s not a viable alternative}.” $T_{5}$ shared that modelers they worked with would often make the argument that “\textit{a .000001 change in the AUC is meaningfully different}.”\footnote{The Area Under the Receiver Operating Characteristic (AUC) curve plots the true positive rate against the false positive rate for all possible thresholds.} We note that the way fair lending teams measured model performance, and what metrics of performance they traded off with disparity, also varied widely, with some using AUC,  others using KS, others simply using accuracy, and still others forecasting directly to dollars lost.

Notably, participants ($T_{5}$, $T_{9}$) observed a level of internal inconsistency between model degradation tolerated for other model risk management processes and for fair lending, which they thought was unjust. As $T_{9}$ said: “\textit{If you allow your model, before you have to} [retrain to deal with drift]\textit{, to degrade by 2 and a half percent, 5 percent, why can't you allow that} [to combat disparate impact]? \textit{If it's not meaningful for you from a safety and soundness standpoint, why is it meaningful for you from a }[disparate impact] \textit{standpoint}?” Some participants suggested some of this conflict could be the result of ``\textit{cultural issues}'' with data scientists, stemming from how they  ``\textit{were trained.'' ($T_{8})$.} \looseness=-1

However, most participants believed that the lack of uniformity here was in part due to a lack of regulatory clarity around an acceptable threshold---and some specifically noted they were eager for one
($E_{11}$, $L_{6}$, $E_{8}$, $L_{7}$). As $E_{8}$
shared, “\textit{My job will be easier if tomorrow the CFPB comes with prescriptive guidance [...] with a number}.” However, as we see in Section~\ref{sec:find:no_consistent:tradeoff}, finding the line between clarity and specificity can be a difficult task.
And, as we discuss in the next section, clarity in regulatory expectations can only advance work so far when perceived legal tensions linger.

\subsubsection{Perceived tensions: disparate treatment and impact}
\label{sec:find:no_consistent:tensions}
Fair lending programs must manage risks related to two doctrines in discrimination law, disparate treatment and disparate impact: while disparate treatment outlaws decision-making \textit{on the basis of protected characteristics} such as race, disparate impact aims to reduce unjustified 
disparities in decision-making outcomes. Participants most often mentioned challenges navigating both theories simultaneously when discussing how their organizations think about mitigating gaps revealed by disparate impact analyses, particularly with regard to the extent to which demographic information may be considered when searching for LDAs.

A discussion of the legality of algorithmic bias mitigation techniques in light of disparate treatment constraints is out of scope of this paper (for a detailed treatment of this issue, see \cite{kim2022}). Nonetheless, our research underscored the practical impact of the \textit{perception} that a direct tension between these two legal doctrines exists, especially in the case of AI systems, within the fair lending community. Participants' reflections revealed how this perceived tension influenced companies' bias mitigation procedures, which they observed regulators have generally found to be acceptable. However, even as many practitioners find these procedural design choices to be increasingly outdated in light of more complex AI systems, they observe that their firms have been reticent to update approaches without an explicit go-ahead from the regulator. At the same time, regulators were aware that while they could use their discretion to communicate expectations, ultimately they alone cannot resolve this sort of contentious topic.

\noindent\paragraph{Participant frustration with seemingly inefficient institutional design choices.} 
One popular strategy financial institutions employ to navigate the perceived tension between disparate treatment and impact is a complete separation of individuals working on building financial models subject to fair lending law (the “first line of defense” in firms' risk management structure) and those investigating them for, and potentially mitigating, bias (the “second line”). 
This separation typically dictated access to demographic data: many participants shared that only the second line would have access to demographic data necessary to conduct fairness tests. $E_{4}$ explained, ``\textit{
you never gave the modelers access to the fair lending attributes 
[since you]
...didn't want anybody putting race or ethnicity, or anything like that in a model. 
}''

Many participants confirmed this set-up 
($E_{1}$, $E_{4}$, $E_{6}$, $L_{6}$, $T_{9}$): for example, $E_{4}$
noted that \textit{“modelers did not gain access to protected class data.}” 
This firewall also prevented first line modelers from even seeing the results of the testing done by the second line 
($E_{1}$, $E_{6}$, $E_{8}$): For example, $E_{6}$
observed that if the first line “\textit{got a negative outcome from the disparate impact analysis},'' they weren't ``\textit{allowed to see}''  the disparate impact analysis “\textit{so they would just be told to change the model, but then} [they] \textit{wouldn't know how to change it}.” 

While this organizational practice may have been a plausible way to manage discrimination risks in a paradigm of simpler, smaller models for which new versions could be developed with reasonable ease, participants shared that such structures serve as a serious impediment to effective mitigation when dealing with today's more complex, larger models 
($E_{1}$, $E_{6}$, $L_{6}$, $T_{9}$).
As a result of the separation and lack of communication between the two teams, $E_{6}$ described an inefficient ``\textit{back and forth, trying to figure out how to fix the model without having access to analysis}.'' $T_{9}$ agreed, stating that the first line/second line separation meant that “\textit{you need to have two complete new model teams. It just doesn't really work well.}” $L_{6}$ expressed that in “\textit{an ideal world},” the first line would do the bias testing alongside the model development, since explaining how the model works well enough to another team for them to test it takes "\textit{a lot of time}." 
Several participants expressed deep frustration with this set-up, some even leaving their jobs over the practice. As $E_{1}$ explained, ``\textit{[D]on't you want our data scientists to understand fairness so that we are engineering it in at the point of model development rather than doing it after the fact? And so, it took a very long time for} [the second line] \textit{to even warm up to the idea that they should even talk to me as representing part of the business. This is something that  honestly I was not able to successfully move the needle on very much at }[company name],\textit{ which is also part of the reason why I left.}''

Some participants observed their companies starting to deviate from this structure 
($E_{11}$, $L_{6}$, $T_{9}$). As $L_{6}$
noted, some firms have developed a process where the first line develops the best performing model, “draws a line,” and then performs disparate impact analysis. That is, they separate the two processes within one team, to preserve independence of the processes while reducing information flow issues. $T_{9}$ agreed that firms are moving to establish a “\textit{clear delineation as to who in the first line is going to do the }[disparate impact] \textit{work [...] I think that’s where we’re all kind of landing right now}.” 

However, progress towards this change appears to be slow, and marked by continued concerns over implicating disparate treatment. For example, $L_{7}$ described the process at their institution, where first line developers did have access to some basic disparate impact information through an “\textit{automatic bias estimator [...] where the modelers can actually do a little bit of} [disparate impact analysis] \textit{during the development process without necessarily giving the modelers access to race and ethnicity data}.” Critically, though, $L_{7}$ noted that their organization places limits on how many times modelers can use the automatic bias estimator: “\textit{You could in theory [...] run iterations through 150 different times, and basically snuff out [...] it sort of becomes a backhanded way of getting to the demographic information},'' and that in response they impose a limit on ``\textit{how many runs you can put through the thing, to specifically avoid that problem}.” 
Some viewed such approaches as a practical path forward, allowing LDAs to be surfaced earlier in model development while mitigating disparate treatment concerns by not revealing raw demographic information to models or modelers.

\paragraph{Frustration with reliance on out-of-date mitigation methods.} 
Participants described that, in practice, many methods that have been proposed by researchers to mitigate discrimination in AI systems are seen as inducing a risk of disparate treatment discrimination~\cite{ho2020affirmative, kim2022}.\footnote{Following \cite{kim2022}, we disagree with this view, but do not address it here as it is out of scope for this work.} This seemed to be driven by firms' interpretations of disparate treatment---which prohibits decision-making directly on the basis of a protected attribute such as race---to mean that protected class information should not be used at all during the course of model development. As $L_{5}$ shared, “\textit{Those of us in the industry don't think we can actually factor a person's demographics into [...] the building of a model in order to lessen discrimination, at least not factored in directly [...] Whereas there are [...] academics or others outside of the industry that feel like that is a way mathematically that can be helpful}.” However, much of the algorithmic fairness literature assumes at least some access to demographic data, and model debiasing methods informed by demographic information are understood to be more effective at reducing algorithmic discrimination than anticlassification  (i.e. fairness through unawareness).
Perceived legal risk has led many financial institutions to opt not to use 
 these more modern debiasing methods.
 As $T_{9}$ explained, while regulators in recent years expressed an expectation that firms be proactive in mitigating bias, they “\textit{haven't fully signed off on}” many algorithmic debiasing methods financial institutions may want to use to reduce disparate impact. $E_{6}$ also stated that their fair lending department was resistant to deviating from the status quo by amending legacy methods: when they tried to collaborate with their firm's fair lending team, $E_{6}$ found that fair lending insisted that their “\textit{methods are good enough, and they} [fair lending] \textit{get very defensive [...] the paradox is that you can't improve procedures if your position is that legally, we are doing everything we need to do}.” Indeed, for many participants, basic methods like drop-one analysis were the main or only debiasing method discussed during our interviews 
($T_{2}$, $R_{1}$, $T_{9}$, $E_{9}$).
 $R_{1}$, a former regulator, affirmed that in general, 
 drop-one or variable changes ``\textit{tends to be as far as companies go}.”

Despite this inertia, participants shared an overall recognition that such approaches were no longer effective for more complicated AI systems. Whether or not their firms permitted use of newer methods from the algorithmic fairness literature, most of our participants believed that such methods would be necessary to actually tackle biases in AI systems 
($E_{4}$, $E_{6}$, $E_{11}$, $E_{9}$). For example, $E_{11}$
stated that “\textit{drop one is an old school method that that doesn't work anymore}.” Thus, if a company is not willing to experiment with newer methods due to fears of implicating disparate treatment---which many were not---our data suggests that this perceived tension led to a much less effective bias mitigation approach. 

\subsubsection{Navigating uncertainty around regulatory expectations, and its consequences}
\label{sec:find:no_consistent:tradeoff}

As shown in Section~\ref{sec:find:floor}, while fair lending practitioners understood they needed to test their systems for discrimination and search for LDAs, they also had little concrete guidance on how to do so. This lack of clarity resulted in a significant variety of processes.
As $E_{11}$ stated: 
``\textit{I'm not even sure I've seen two institutions do bias testing and LDA (especially allowable performance loss) exactly the same way; it's kind of wild how different it is.}''

Some participants reported this lack of clarity afforded them room to experiment with new and, in their experience, more effective methods and structures to reduce discriminatory outcomes. As discussed above in Section~\ref{sec:find:no_consistent:tensions}, participants observed their firms' propensity to stick to established yet inadequate methods, but at the time of our data collection many were beginning to reconsider them ($E_{1}$, $E_{4}$, 
$T_{5}$, $E_{9}$). $E_{9}$ shared that “\textit{as we use machine learning models [...] in the industry, everybody is moving away from a manual review }[i.e. drop one], \textit{because that cannot suffice}.” 

Indeed, some participants reported that their firms interpreted the lack of explicit regulatory guidance as an opening to explore different procedural structures to allow some awareness of demographic trends in aggregate while still managing disparate treatment risk. 
For instance, several participants expressed comfort with hyperparameter-tuning or “pipeline aware” methods ~\cite{black2024less} to search the design space of models, using demographic measurements only to choose among models after training 
($E_{1}$, $E_{4}$, $T_{5}$). $E_{1}$
described their firm's LDA search process, explaining that they would \textit{ “search the entire design space. But obviously, for practical reasons, we're going to spend only X amount of dollars to search Y number of alternative models, and with the Y number of alternative models that we have done some sort of like semi-exhaustive search...that fulfills our need to show that we could not have done any better.”} $E_{1}$ continued, noting that using demographic data as an iterative reward to search the design space seemed not to bring up disparate treatment concerns since it was simply used to measure reward after experimenting with feature swapping, feature engineering, or hyperparameter tuning: "[Legal and compliance] \textit{were happy with that level of arm's length where we're not explicitly interfering [...] they were happy to sign off that this is a good methodology for this purpose.}” $E_{4}$, also noted that they had adopted more modern debiasing methods: ``\textit{we relied on generative adversarial networks} [for debiasing...] \textit{that was way more impactful than the old drop-one analysis}.''
Had regulators preordained a suite of methods, their guidance may not have been capacious enough to accommodate such exploration, further disincentivizing proactive fairness testing.

At the same time, other participants reported how this same regulatory ambiguity incentivized their firms to do the bare minimum, served as a challenge or blocker to fair lending work, and ended up orienting fair lending teams' work mainly to document justifications of unaddressed disparate impact that they feel the regulator would accept rather than actually mitigating disparities. Importantly, existing disparate impact doctrine does not \textit{require} firms to extinguish any disparate impact against a protected group. Instead, the reach of disparate impact doctrine is more limited: it merely requires firms to establish a business justification for the practice that results in a disparate impact. Specifically, under the Equal Credit Opportunity Act (ECOA), a financial institution must show that its practice ``meets a legitimate business need.'' As a result, $E_{11}$ described that “\textit{many institutions want to have a credible performance of doing the LDA search because they're mostly concerned with being able to display something that satisfies the regulator.}” But $E_{11}$ noted that \textit{“there is a large gap between ‘we need to demonstrate that we're trying here' versus 'we actually want to do an LDA search that finds better alternatives'.}"

In stark contrast to $E_{1}$'s firm's thorough search described above, $L_{5}$'s firm often didn’t even test models subject to fair lending law because they were confident they could justify typical models as simply common practice: “\textit{For a linear or logistic regression, or even some of the AI techniques 
[...] if the inputs are basically your credit data, or what is normally collected on an application, as a rule, we don't feel like there's there's real value in doing the statistical testing, in part, because we're not sure what we would change [...] mostly because the law and the regulatory expectations are actually very unclear.}” Regardless, many participants experienced how fair lending work can end up an exercise in defensive documentation---that is, justifying the legality of an ongoing practice to a regulator, rather than earnestly searching for less discriminatory decision-making systems. In fact, such “justificatory work” appeared to be core to what financial regulators have expected to see in the past, and core to what fair lending programs are tasked to produce. One example that participants described is a firm’s business justification memo, report, or slide deck---an artifact that is often prepared by financial institutions explicitly for a regulatory audience if a model the firm wants to launch has been determined to exhibit a disparate impact. According to $T_{9}$, this document would contain a description of the model, a list of variables, the performance of the model, results of tests conducted, whether or not any viable LDAs were found, and “\textit{if there were, why they weren't taken, or why they were implemented}.”

Some institutions were described to be very comfortable with justifying disparities using projected profit losses, or by arguing that the practices leading to them were industry standard enough to justify the disparity without even looking for other models ($L_{5}$, $E_{9}$). $E_{9}$ explained as an example that ``\textit{Blacks could have a lower FICO score than Whites. But that's because the actual default risk that has been observed for the Black class is higher. So that's one business justification}," among other factors a firm could use to explain away remaining disparities. Another way participants described their firms defining rules that inhibited the effective prevention of discrimination was to allow for zero performance drop when searching for a less discriminatory alternative. $R_{1}$ noted that several companies were "\textit{0 change}"---that is, tolerated no change in performance metrics for an LDA, and justified that position with their profit interest. They queried whether that justification was truly sufficient, but conceded that such was the act of businesses conducting risk management.

\section{Discussion}
\label{sec:discussion}
\subsection{More mature practices are still plagued by familiar challenges}
\label{disc:plague}

While financial institutions' bias mitigation practices appear far more mature than those of 
entities not covered, 
financial practitioners still face many of the same problems that have surfaced in entities not directly covered by civil rights law \cite{madaio2020co, smith2025pragmatic, cinca2025practitioners, akbar2024trustworthy, costanza2022audits, schiff2024emergence}. Though the hurdles are familiar, their trajectory in the financial sector reveal notable differences that offer RAI researchers, as well as practitioners in less mature organizations, a glimpse into a potential future.

One shared problem is limited access to demographic data. While entities not covered by fair lending law typically experience this as simply missing or insufficient demographic data for fairness testing purposes~\cite{10.1145/3351095.3372877,andrus2021we}, financial institutions have established methods (e.g., BISG) for race data but still lack methods for most other demographics. Sex and other characteristics still protected by law such as national origin and disability 
are often missing: $T_{10}$ shared that variables such as disability 
were ``\textit{much, much more difficult to collect data on}.'' 
While the use of BISG, despite its recognized limitations, has been accepted by regulators, the existence of such a method at all has created a perverse incentive to avoid 
collecting self-identified race data. 
Thus, financial institutions generally lack it (except for mortgage loans, where applicants 
may disclose this information), even though they have the option to collect it from borrowers ~\cite{ecoa1974}.

Second, since the decision about whether or not to mitigate disparities rests with the business side of the firm, fair lending practitioners still have to foster company and colleague buy-in informally,  similar to RAI practitioners \cite{schiff2024emergence, orr2020attributions, deng2023investigating, deng2023understanding, ali2023walking}, in order to motivate substantive rather than defensive antidiscrimination work.
While research has pointed to gaps in regulation to explain corporate inaction on fairness, we observe that even with the specter of regulatory scrutiny to  motivate fairness work, participants 
($L_{2}$, $T_{2}$, $E_{4}$) described needing to leverage appeals to pathos as a way to successfully lobby their organizations to adopt more proactive fairness approaches, posing questions to colleagues and decisionmakers like 
“\textit{would you really want your grandmother to know that we were doing this}?” ($L_2$). 

Finally, the research-practice gap that challenges ordinary institutions seems particularly acute in the highly regulated financial industry. From a technical perspective, fairness problems in fair lending starkly diverge from those contemplated in the literature: information flow 
between model builders and testers (first line/second line) about measurement results is significantly more limited than research typically envisions. Testing is typically not done with true demographic data, but imputed proxies, implicating measurement challenges that have received 
less research attention than other topics. And, practitioners must consider more than the two or four protected groups typically considered in research---fair lending work must be done over a ``\textit{2 by 7 matrix}'' ($E_1$) of demographics. This can lead to the ``train-crash`` problem, where it is combinatorially challenging to simultaneously reduce disparities for all protected groups ($E_1$, $E_{9}$). While a great deal of theoretical work has addressed this problem~\cite{hebert2018multicalibration,gopalan2021omnipredictors}, practitioner-ready tools have yet to follow.

Beyond these familiar themes, the lack of attention paid in the development of bias mitigation methods to practical tensions between disparate treatment and impact has meant that most 
algorithmic debiasing methods are seen as too legally risky to adopt in practice, even if thorough technical and legal review would suggest they are not \cite{kim2022}. This tension results in a practical ceiling of fair lending efforts at the model level, leaving fair lending practitioners in a position where using inferior methods for bias mitigation is safest, underscoring the importance of basing research agendas on empirical insights from the field.

\subsection{The mixed success of fair lending offers useful lessons for regulatory proposals for AI governance}
\label{sec:disc:mixed}
The history of fair lending offers a cautionary tale for those looking to regulation to solve the many ills of algorithmic decision making. While our empirical findings support the intuition that  regulation operationalized through vigorous and proactive enforcement mechanisms can foster desirable practices that firms might not adopt voluntarily, they also highlight that the success of even such potent regulation is not a sure thing. Developing effective regulation is a far more challenging task than the current literature's countless policy proposals might seem to suggest.\looseness=-1

To begin, regulation is commonly developed without sufficient attention to administrability---that is, how firms will attempt (or not attempt) to comply with the law in practice \cite{GuhaEtAl2024AIRegulationAlignmentProblem}. Fully anticipating, ex ante, how firms will respond to regulation---as well as the organizational and practical challenges that they will face in seeking to comply with regulation---is genuinely challenging. Even recognizing these challenges, regulatory proposals are often concerningly detached from the realities of compliance. It's not enough for regulation to have ``teeth'' in theory; effective regulation takes into account how compliance will work in practice. This may involve accounting for natural disincentives to comply fully, with commensurate penalties to shift firms' risk calculus. In other cases, it may require understanding the everyday challenges that practitioners face even when they are sufficiently incentivized to attempt to comply.

Second, regulatory proposals are too rarely informed by evidence demonstrating the relative efficacy of different policy interventions. While there have been recent calls to make AI policymaking more evidence-based (e.g., \cite{BommasaniEtAl2025AdvancingEvidenceBasedAIPolicy}), these calls tend to focus on evidence about the underlying risks that regulation aims to address rather than about the efficacy of different regulatory designs \cite{baldwin2011understanding, cohen2024regulatory}. Likewise, while there have been recent efforts to evaluate how well regulations targeting algorithmic systems actually work in practice (e.g., \cite{10.1145/3630106.3658998,10.1145/3630106.3658959,10.1145/3715275.3732004, wright2024nullcompliance}), these evaluations rarely engage with questions of what regulatory designs could lead policy interventions to be more effective. Researchers looking to channel their work toward policy recommendations that seek to address the challenges of algorithmic systems should expand their analysis beyond the substance of potential policy requirements and more actively consider the relative successes of different regulatory designs in operationalizing relevant policy goals.

In particular, it is striking how minimally the decades of experience around compliance with and enforcement of fair lending laws have influenced regulatory debates about AI at a structural level.\footnote{One potential reason that the decades of fair lending experience has been overlooked is that fair lending supervision is purposefully secretive and confidential. This itself is a trade-off in regulatory design. While such secrecy can empower regulators and regulated entities to share information more candidly, and can even cultivate innovative practices, the secrecy can, absent significant external research, essentially prevent advocates, researchers, and policymakers from understanding the on-the-ground regulatory reality or lock in industry-preferred but less effective practices.} 
For example, our findings highlight the unique role that supervisory and enforcement authority have played in successfully fostering fair lending practices---a regulatory design feature that is distinct from other areas of civil rights law and almost completely absent from recent policy proposals for dealing with algorithmic discrimination. Future research could help to avoid such oversight by helping ensure policy interventions and their regulatory designs are directly informed by the lessons learned from empirical studies of compliance, both in historically regulated domains like lending and in the more recent area of responsible AI. 
To date, the work advancing policy proposals for AI has been largely divorced from the work investigating the problems that AI policy interventions face in practice \cite{GuhaEtAl2024AIRegulationAlignmentProblem}. While each of these areas of research have reached a point of saturation (the literature on AI policy has ballooned in recent years, while empirical studies of responsible AI in practice have begun to report overlapping findings), there is enormous potential for work that marries the two. More research into the everyday work of enforcement and compliance will go a long way toward ensuring that efforts to craft both substantively and structurally effective regulations are best positioned to succeed. 

\subsection{Discretion is a double-edged sword}
\label{sec:disc:design} 
Clearly, the regulatory design of fair lending supervision is an area ripe for further exploration by researchers, advocates, and policymakers interested in holding discriminatory AI systems to account. As the findings in this paper suggest, robust supervision and examination can create the conditions for routine, ongoing interventions that fundamentally shape companies’ antidiscrimination efforts. While the efforts described by our findings were insufficient in fully combating discrimination, they do point to a potential regulatory path forward.

Nevertheless, regulatory design fundamentally dependent upon the exercise of significant institutional discretion---in particular, discretion as exercised within supervision---can have significant drawbacks. Discretion is not a one-way ratchet toward a static policy goal. In fact, the opposite can be true: discretion is inherently downstream of political ends that may quickly change between administrations. For example, before the CFPB took formal regulatory steps to significantly amend Regulation B and eliminate disparate impact liability \cite{cfpb2025ecoa_fr}, and even before the Trump administration promulgated an Executive Order that suggested disparate impact liability “violates our Constitution” \cite{eo14173}, the CFPB merely noted in a memo that they “will not engage in redlining or bias assessment supervisions or enforcement based solely on statistical evidence” \cite{cfpb2025ecoa_fr}, essentially ending review of potential disparate impact. Such a move highlights how potentially unstable regulatory regimes predicated on the exercise of significant discretion in supervision can be. 

\section{Limitations} 
While our sample includes 35 participants, it is not representative of all fair lending practitioners. For example, we did not interview individuals from smaller financial institutions or state regulators. Legal sensitivities and social desirability bias~\cite{braun2021thematic} may have also shaped participants’ accounts, leading to incomplete 
descriptions. Further, 
this study 
reflects views that have likely evolved since data collection, especially given changes in the federal regulatory environment. Finally, given the lack of prior work, our goal is to offer an empirical account of fair lending practices as they exist on the ground and the factors shaping them, rather than a critical analysis. Future work can and should provide critique.

\section{Conclusion} 
For years, the FAccT community has called for upstream, proactive testing of models that power high-stakes automated decision systems. As our findings show, the regime of fair lending law helped create these very processes, albeit under unfamiliar monikers. Whether readers view these processes as unduly narrow and 
examples of fairness washing, or exemplars of what ongoing, robust supervision can accomplish, we believe that the multi-decade effort to implement fair lending protections holds important lessons for any effort to combat algorithmic discrimination. Regardless, civil rights law is under attack. 
Merely reviving previous  protections will require concerted effort. In parallel, policymakers are proposing AI-specific interventions to challenge discrimination without the benefit of critical insights from the trenches of fair lending compliance. We believe that both such efforts to strengthen civil rights protections can find value in grappling with the successes and limitations revealed by our results.

\newpage

\section{Generative AI Statement}
The authors used generative AI to help format bibtex citations of references selected by the authors. There was no other use of Generative AI assistance in making this manuscript. 
\bibliographystyle{ACM-Reference-Format}
\bibliography{bib, industryRAIpractice}

\appendix
\section{Participant Information}
In this section, we include our participant information in Table~\ref{tab:participants}.

\begin{table}[h]
\centering
\textbf{Table 1}
\vspace{0.3cm}

\begin{tabular}{|c|l|c|l|c|l|}
\hline
Participant & Population & Participant & Population & Participant & Population \\
\hline
P1 (L1)  & Lawyer                 & P15 (E6)& Engineer                & P29 (T10) & Third Party \\
P2 (L2)  & Lawyer                 & P16 (T6) & Third Party             & P30 (E8)& Engineer \\
P3 (T1)  & Regulator, Third Party  & P17 (T7)  & Third Party             & P31 (E9) & Engineer \\
P4 (T2) & Third Party             & P18 (R1) & Regulator, Engineer     & P32 (R7) & Regulator \\
P5 (L3) & Lawyer                 & P19 (R2) & Regulator               & P33 (L7) & Lawyer \\
P6  (E1) & Engineer               & P20 (R3) & Regulator               & P34 (E10) & Engineer \\
P7  (E2) & Engineer               & P21 (R4) & Regulator               & P35 (R8) & Regulator \\
P8 (E3) & Engineer               & P22 (T8) & Third Party             & & \\
P9 (L4) & Lawyer                 & P23 (E11) & Regulator, Engineer     & & \\
P10 (T3) & Third Party             & P24 (L5) & Lawyer                 & &  \\
P11 (T4) & Third Party             & P25 (L6)  & Lawyer                 & &  \\
P12 (E4) & Engineer               & P26 (T9) & Third Party, Lawyer     &  & \\
P13 (T5) & Third Party             & P27 (R6) & Regulator              &   & \\
P14 (E5) & Engineer               & P28 (E7) & Engineer                &     &  \\
\hline
\end{tabular}
\caption{Participants' professional backgrounds. Following prior work on responsible AI practices in industry \cite{deng2023investigating, madaio2024tinker, widder2024power}, we omitted demographic details and abstracted certain information about participants’ employers and roles to avoid inadvertently identifying individuals working at the forefront of this high-stakes domain.}
\label{tab:participants}
\end{table}

\section{Interview Protocols}
The following section provides the protocols used in our semi-structured interviews with policymakers, engineers, and regulators. Note that due to the nature of semi-structured interviews, some questions in the protocol were omitted and follow-up questions were asked instead.
\subsection{Policymakers}
\label{app:protocol_policy}

\vspace{2mm}
\noindent \textbf{Background, process, and organizational structure}

\noindent \textit{In this interview, we’re interested in the process by which models are trained and evaluated, especially with regards to fairness, and whether and how legal and compliance interacts with these processes.  We appreciate that you’re not able to share specific legal advice or information protected by legal privilege and just to emphasize we aren’t looking for you to share that information, we are interested in more general concepts ; feel free to phrase your answers in the hypothetical or in general in order to answer to whatever degree of detail you are able to.}

\noindent \textbf{Q1} Could you tell us briefly about what your role is at your current organization and what brought you to this type of work? 

\noindent \textbf{Q2} We are talking with people at companies across different contexts, for example financial services, online platforms, housing, and more. What application space is your team or company working on?

\noindent \textit{We’re interested in learning about different processes different firms have related to algorithmic fairness or discrimination. First, we’ll be asking about your firm’s approach.}

\noindent \textbf{Q1} Is there a structured process by which teams at your organization are performing work related to algorithmic fairness or discrimination of any kind?
\begin{enumerate}
    \itemsep0em
    \item Does your organization perform any tests for algorithmic discrimination, for example testing for disparities in model behavior across demographic groups? Why or why not?
    \begin{enumerate}
        \itemsep 0em
        \item If so, talk us through that testing process. What triggers the testing process to start?
        \item When are tests performed, and what tests are done?
        \item What, if any, is your firm’s policy in the case that a disparity is discovered?
        \item  When, if ever, would your firm look for less discriminatory models to replace a potentially discriminatory model? How would this search take place?
    \end{enumerate}
\end{enumerate}

\noindent \textbf{Q2} Does your organization have defined policies, practices, or processes for decision making when conducting work on AI fairness or discrimination? For example, does your organization have written policies and procedures about how to evaluate and mitigate disparities or search for fairer models?
\begin{enumerate}
    \itemsep0em
    \item Do you know how your organization came to these policies and processes?
    \item Does your organization have any norms or explicit policies around documenting algorithmic fairness efforts?
\end{enumerate}

\noindent \textit{For the next few questions, we’re curious about your firm’s organizational structure around fairness testing.}

\noindent \textbf{Q1} Is there a fairness/anti-discrimination/fair lending team? If so, how is that team structured? What is the relationship between this team and other teams, including model developers? 
\begin{enumerate}
    \item If so, how is that team structured?
    \begin{enumerate}
        \item What is the relationship between this team and other teams, including model developers? 
        \item What role does this team play: are they a gate, guide, or partner?
        \item Who is responsible for implementing the team’s recommendations?
        \item Whom does that team report to?
        \item Do you know why the work was organized in this way? For example, were there any legal considerations?
    \end{enumerate}
    \item If not, where does fairness work occur in the organization?
    \begin{enumerate}
        \item What role do those who do fairness work play– guides, gates, or partners?
        \item Who do the fairness practitioners report to?
    \end{enumerate}
    \item How are you personally involved in this work? Where do you come in in this structure?
\end{enumerate}

\noindent \textbf{Q2} Who makes key decisions about this work? For instance, setting priorities, methods,  policies, etc. 
\begin{enumerate}
    \item Who is responsible for making decisions about whether or not a given model is deployed? 
    \item Are you involved in any decisions for what measurements or mitigations are necessary, or which models are ultimately appropriate to deploy?
    \begin{enumerate}
        \item If yes:  What factors tend to inform the recommendations you would generally make in these sorts of conversations?
        \item If no: Can you give me your sense of what goes into these decisions?
    \end{enumerate}
\end{enumerate}

\noindent \textbf{Q3} Given that organizational structure, how do these teams (either fairness/fair lending/ or individuals who do fairness work outside of a team)  interact with other teams at your organization — in particular ML/AI developers  — to conduct fairness work?
\begin{enumerate}
    \item What is the relationship between these teams like?
    \item Are there any things you’ve observed to be particularly challenging when less-technical stakeholders communicate with technical colleagues as they do this work?
    \item Are there any other challenges you’ve experienced that are relevant to fairness work at your organization? 
\end{enumerate}

\noindent \textbf{Q4} Are there any internal policies that seem to constrain your organization’s ability to test and mitigate algorithmic disparities?
\begin{enumerate}
    \item How do you think your organization's process as it stands could be improved to fulfill its higher order goals of preventing discrimination?
    \item What do you think would need to change in order to accomplish these goals?
\end{enumerate}

\vspace{2mm}
\noindent \textbf{Laws that underwrite the process (DI) and interpretations/ regulatory expectations}

\noindent \textit{As we mentioned, we’re interested in how models are evaluated and trained, especially with regards to fairness, and whether and how legal and compliance interacts with these processes. For the next few questions, we’re interested in how laws, regulations, and policies shape these efforts.}

\noindent \textbf{Q1} Disparate impact has informed a huge range of algorithmic fairness/fair lending/anti-discrimination work — how do you think that disparate impact doctrine relates to your work? Does it inform the work your firm does around fairness testing?
\begin{enumerate}
    \item If so, tell us a bit more about what you believe it requires your organization to do, or not do?
    \begin{enumerate}
    \item How does the doctrine relate to the process by which your firm tests for algorithmic discrimination?
    \begin{enumerate}
        \item How does the doctrine relate to searches for fairer models? 
        \item Do different models or model types fall under or outside of the DI framework at your organization? What products or models don’t? What makes these products or models different from one another? For example, is there an explicit classification of low risk/high risk products? How are models classified into each?
    \end{enumerate}
    \item How does your team think about the second step of the disparate impact test — the business justification of a model?
    \end{enumerate}
    \item If not, do you have a sense of why this doctrine is not seen to directly apply to your business?
    \item <If they have a fairness testing/mitigation process> What motivated your company to implement discrimination testing (and mitigation), given the lack of a legal requirement?
    \begin{enumerate}
        \item In absence of a legal framework, are there any principles your firm uses to guide the testing and mitigation of algorithmic bias? For example, are there any particular factors your organization considers with regard to the defensibility of launching a product that may have a disparate impact or launching a particular version of a model over a different version?
    \end{enumerate}
\end{enumerate}
\noindent \textbf{Q2} In your opinion, are there scenarios where it would be clear that a team like yours, or an organization like yours, would be required to search for an LDA? 
\begin{enumerate}
    \item If yes, does that happen today? 
    \item If no, can you imagine any?
    \item In thinking about LDAs that serve the same business purpose, what does ``serve the same business purpose'' mean to you? Can you talk about what that concept  means to you in your work?
\end{enumerate}

\noindent \textbf{Q3} Are there laws, regulations or guidance that seem to constrain your organization’s ability to test models for potential performance differences across demographic groups?
\begin{enumerate}
    \item If you could make changes to such laws, what would they be and why?
\end{enumerate}

\noindent \textbf{Q4} Do you believe that regulatory agencies (such as, e.g. the CFPB/EEOC/FHA) have certain expectations of how your business performs fairness tests and/or bias mitigation? Why or why not?
\begin{enumerate}
    \item What do you perceive those expectations to be, if so?
    \item For the relevant regulatory agency in your domain (e.g. CFPB, HUD, FTC), how do you keep up to date on any guidance whether formal or informal?
\end{enumerate}

\noindent \textbf{Q5} Do you think your organization, or organizations like yours, would benefit from more regulatory clarity or guidance on how to search for LDAs? What would that look like, hypothetically?

\vspace{2mm}
\noindent \textbf{Cost and other constraints}

\noindent \textbf{Q1} At organizations like yours, to what extent are costs actively considered when determining what models should be tested? 
\begin{enumerate}
    \item How about the costs involved in searching for fairer models (LDAs), or what model to choose after searching? 
    \item What are some of the costs that you’re aware of that play a role in decision-making?
\end{enumerate}

\noindent \textbf{Q2} Are some more important than others?
\begin{enumerate}
    \item How does your organization navigate these costs in the context of compliance?
    \item Is this consideration explicit or subtextual?
\end{enumerate}

\noindent \textbf{Q3} Are there other factors beyond cost that are considered when determining which models to pursue or to launch?
\begin{enumerate}
    \item Can you share about what sort of factors are considered, and to what extent?
    \item Do any particular factors tend to trump others?
\end{enumerate}

\vspace{2mm}
\noindent \textbf{Wrap up}

\noindent \textbf{Q1} Is there anything you wanted to say that you didn’t get the chance to?

\noindent \textbf{Q2} Is there anyone else you could connect us with that might be a good fit for our interview?

\subsection{Engineers}
\label{app:protocol_engineer}

\vspace{2mm}
\noindent \textbf{Background and overall process}

\noindent \textbf{Q1} Could you describe what application space your team is working on?

\noindent \textbf{Q2} Does your organization have written policies and procedures about how to evaluate your models? How about how to mitigate disparities or search for fairer models?
\begin{enumerate}
    \item Is there a designated team who’s responsible for this? Are you a part of this team?
    \begin{enumerate}
        \item What is the relationship between this team and your team?
        \item Who do you report to?
        \item And who does the fairness team report to?
    \end{enumerate}
    \item Do you have legal and compliance teams? How big are the teams? Who do they report to?
    \item Do you have an inventory of all your models?
\end{enumerate}

\noindent \textbf{Q3} Imagine you’re deploying a new model in your team <insert appropriate concrete example>. Can you briefly take me through the process of testing, reviewing, and deploying that new model?
\begin{enumerate}
    \item What would be the decision points about specifically what model versions to launch? 
    \item Do you do any testing of your model or model versions across different populations?
    \begin{enumerate}
        \item What are the different groups? Do they have any relationship to demographic groups? 
        \begin{enumerate}
            \item Which demographic groups and why?
            \item How do you measure demographic groups?
        \end{enumerate}
        \item Under what conditions do you do this testing? 
        \item How broad is the testing?
        \begin{enumerate}
            \item How much time does it take and how many people are involved?
            \item Do you have a sense of how much it costs the organization?
        \end{enumerate}
    \end{enumerate}
    \item Does your model building process involve proactively testing for differences across demographic groups, or does it do so in an ad-hoc fashion? Or, does it just happen at the end?
    \begin{enumerate}
        \item Does your team perform these tests or is there some other team that does? 
    \end{enumerate}
\end{enumerate}
\noindent \textbf{Q4} <If they test for fairness problems> How did you learn what you should be looking out for when performing fairness/bias assessments?
\begin{enumerate}
    \item What are the sources of authority on how analyses are done?
    \item Do the recommended tools to search for bias change over time? 
    \item Do you keep up with scholarship on where to look for bias? 
\end{enumerate}

\vspace{2mm}
\noindent \textbf{Metrics}

\noindent \textit{Thanks for providing us with the overall process of your model testing! Now we want to dive a bit deeper into the metrics.}

\noindent \textbf{Q1} Are there specific metrics your org/team considers in building models?
\begin{enumerate}
    \item Given the metrics that your team/org considers when determining the quality of a model, what do you think it means for two different model versions to be interchangeable or similarly good?
    \item How similar do models generally need to be for you to have flexibility to choose between model versions, for example to preference models with more equal behavior across groups
    \item When you’re deciding to deploy a model, what are the most important metrics to perform well on? 
    \begin{enumerate}
        \item Are there ranges of values of those metrics, or a threshold?
        \item Do you consider any counter-metrics? 
        \item Are there any particular metrics which are considered to be business-critical?
        \item Do you ever consider financial factors as a counter metric? If so, what factors?
    \end{enumerate}
\end{enumerate}

\vspace{2mm}
\noindent \textbf{Model comparison}

\textit{Now we’ve talked about the metrics, let’s discuss more about the concrete model comparison based on these metrics}

\noindent \textbf{Q1} To what extent does your organization have infrastructure to systematically compare different model versions?

\noindent \textbf{Q2} Do you have any decision-making process in place for how to choose between model versions?
\begin{enumerate}
    \item In discussions that are working through decision-making processes, are there ever conversations that are weighing general model performance versus performance across groups? 
    \item Are there justifications for differences across groups discussed as a part of that process? 
    \item Who ultimately decides if the justifications are reasonable?
\end{enumerate}

\noindent \textbf{Q3} If your team finds disparities as a result of the testing, what are the next steps?

\vspace{2mm}
\noindent \textbf{Addressing more fair models (technical)}
\textit{ So far, we’ve been talking about testing. Now we want to shift the conversation to how you and your organization go about addressing your findings or proactively searching for alternative models or approaches. <If they offer no words for this themselves> We’re generally interested in the practice of mitigating disparities found in models to be deployed, or finding equally performant models with fewer disparities across predefined demographic groups.  Let’s call those more fair models, or MFMs. Imagine looking for MFMs in your team.}

\noindent \textbf{Q1} When looking for MFMs, do you have a baseline model that you try to mitigate or fix? Or, do you generate and compare a wide range of models?
\begin{enumerate}
    \item Can you describe what this process looks like in practice (for baseline or generating a large set of models)? Are there any standard methods you do (or don’t) use to mitigate disparities / generate more fair models?
    \item <If generating a large set of models> Do you try looking for MFMs in each or multiple model building steps, or do you only look into it at the end once the model is made?
    \begin{enumerate}
        \item If so, how?
        \item What technical infrastructure do you have to search for MFMs at each of these steps?
        \item Do you explore various feature sets?
        \item Do you perform hyperparameter tuning with fairness as a metric?
    \end{enumerate}
\end{enumerate}

\vspace{2mm}
\noindent \textbf{Addressing more fair models (organizational)}
\textit{Ok, now that I have a better understanding of the technical details on choosing more MFM, let’s talk more about the organizational components around this topic.}

\noindent \textbf{Q1} If there is a team that’s empowered to search for MFMs, do they have a set of parameters within which they can operate, like a performance or financial budget?
\begin{enumerate}
    \item Are there any reasons that an MFM might not be deployed if found?
\end{enumerate}

\noindent \textbf{Q2} Are there times when you’re required to search for an MFM? Or are there times when you feel constrained from searching for one?
\begin{enumerate}
    \item Have you or your colleagues experienced any barriers to searching for MFMs? 
    \item <If there is a legal/compliance/fairness team that is not them>: What is your relationship like with <legal/compliance/fairness team>? Do you ever interact with them directly? What are some tough points?
\end{enumerate}

\noindent \textbf{Q3} <URL> Here are some costs organizations might consider when searching for new models. Are these a part of your day to day work, and are there budgets for any of them? Or, is there a different cost metric your organization uses?

\noindent \textbf{Q4} Do you have any process in place to document the policies, practices, or decisions you make when searching for MFMs?

\noindent \textbf{Q5} Who, in your mind, are the key decision makers who decide whether a model will go into production? 
\begin{enumerate}
    \item Does this set of stakeholders change under a different set of circumstances– for example when deciding whether to deploy more fair models? 
\end{enumerate}

\vspace{2mm}
\noindent \textbf{Introducing Less Discriminatory Alternatives}
\textit{Great! Now this is the last part of today’s interview, and we want to ask you if you are familiar with the concept of Less Discriminatory Alternatives.}

\noindent \textbf{Q1} Are you familiar with the concept of less discriminatory alternatives?
\begin{enumerate}
    \item Does your organization frame its fairness work around the concept of less discriminatory alternatives?
    \item Does what we’ve talked about so far (with MFMs) relate to how your organization handles a search for LDAs?
    \item If your org does not think about things this way, have there been discussions about whether to adopt such an approach? What’s your hypothesis as to why the organization has not yet used this approach?
    \begin{enumerate}
        \item If organization does, how many times have you seen an LDA replace a baseline model?
    \end{enumerate}
    \item Does this process seem to reflect a robust and thorough approach to searching for and mitigating disparities, or does it seem like a narrow and compliance-based approach?
\end{enumerate}

\vspace{2mm}
\noindent \textbf{Wrap up}

\noindent \textbf{Q1} Is there anything you wanted to say that you didn’t get the chance to?

\noindent \textbf{Q2} Is there anyone else you could connect us with that might be a good fit for our interview?
\subsection{Regulators}
\label{app:protocol_regulator}

\vspace{2mm}
\noindent \textbf{Background and overall process}

\noindent \textit{Thanks again for doing the study. We appreciate that you are speaking in your personal capacity, nothing reflects the views of the regulator or agencies that you have worked for or currently work for. And just to emphasize: we aren’t looking for you to discuss any specific investigations or names during our conversation, our questions are aimed at more general concepts. For context, we are doing interviews with people who occupy different roles working with algorithmic fairness, such as engineers who design these systems, product managers who help deploy them, policy and compliance managers who help set policy and practice at companies, and regulators such as yourself. We’d like to start with some basic background questions.}

\noindent \textbf{Q1} Could you tell us briefly how you came to work at <agency> and what you do/did?

\noindent \textbf{Q2} Given that background, do you have any experience in your capacity at your current agency or previous agencies with algorithmic discrimination? 
\begin{enumerate}
    \item If yes, please elaborate more, perhaps telling us how you’ve approached the issue.
    \item If no, what about experience with discrimination at <agency>?
    \item Could you at a high-level talk about your agency’s overall approach on discrimination in algorithmic systems? 
\end{enumerate}

\vspace{2mm}
\noindent \textbf{Experience, expectations, and merging the two}
\textit{[Thanks for that background. We’d like to learn more about your experiences with covered entities in relation to LDAs. For the next few questions, we’re really focused on what you have seen companies do in practice.}
\noindent \textbf{Q1} Practically speaking, when it comes to testing algorithmic systems for discrimination, what have you seen regulated firms do? Can you describe the efforts you have seen in practice? 
\begin{enumerate}
    \item What kinds of disparities were tested for? 
    \item How have you seen or observed firms measure disparities?
    \item What are they doing, if anything, if they identify disparities?
    \begin{enumerate}
        \item Are they explicitly searching for less discriminatory alternatives?
        \item How, exactly, have they performed these searches?
    \end{enumerate}
    \item How commonly are these performed across firms? How frequently does each firm perform these tests?
\end{enumerate}

\noindent \textbf{Q2} What do you think accounts for the differences in firms’ practices that you have observed?
\begin{enumerate}
    \item In your experience, what influences firms’ decisions around conducting tests for discrimination in their algorithmic systems?
    \begin{enumerate}
        \item Do firms have different interpretations about what they are legally obligated to do?
        \item  If interpretations seem to differ, why do you think they differ?
    \end{enumerate}
    \item In your experience, what limits or complicates firms’ efforts to test for discrimination in their algorithmic systems?
    \begin{enumerate}
        \item Were there any cases where firms did not do some element of it but ended up having some kind of explanation that the regulator found to be persuasive or reasonable?
    \end{enumerate}
\end{enumerate}

\noindent \textbf{Q3} In your experience, what do firms do when they identify a disparate impact?
\begin{enumerate}
    \item In particular, how do they assess whether there is a business necessity justification for the disparate impact?
    \item Do they attempt to find a less discriminatory alternative?
    \begin{enumerate}
        \item When do firms feel that it is necessary to search for less discriminatory alternatives?
        \item Do they perform searches for any model or only specific kinds of models?
    \end{enumerate}
\end{enumerate}

\noindent \textbf{Q4} When firms search for less discriminatory models, what have you observed them do in practice?
\begin{enumerate}
    \item Where does your sense come from? Conversations? Regular engagement? 
\end{enumerate}

\noindent \textbf{Q5} In your experience, what limited or complicated firms’ search for a less discriminatory alternative model?
\begin{enumerate}
    \item To what extent are these challenges similar to or distinct from those that may limit or complicate efforts to test for discrimination in algorithmic systems?
\end{enumerate}

\noindent \textit{For the next few questions, we’re really focused on what you expect companies or want companies to be doing.}

\noindent \textbf{Q1} When it comes to testing algorithmic systems for discrimination, what were your agency’s expectations of covered entities?
\begin{enumerate}
    \item What’s the reason for those expectations being the expectations? Do you know why those were the expectations?
    \item How were those expectations communicated to covered entities? Were they? 
\end{enumerate}

\noindent \textbf{Q2} Imagine a firm has identified a potential disparate impact — what would you expect them to do to justify their business practice? 
\begin{enumerate}
    \item What would you expect them to document or provide you?
\end{enumerate}

\noindent \textbf{Q3} When it comes to searching for less discriminatory models, what were your agency’s expectations of covered entities?
\begin{enumerate}
    \item Was the expectation that LDAs need to be equally effective in performance to be considered viable alternatives?
    \begin{enumerate}
        \item If yes, what informs your understanding?
        \item If not, what are the requirements on LDA performance?
    \end{enumerate}
\end{enumerate}

\noindent \textbf{Q4} What guides your agency’s <enforcement/supervision/research> on algorithmic discrimination?
\begin{enumerate}
    \item What was your agency’s investigatory capacity like?
    \item What does it look like to do target selection at your agency?
    \begin{enumerate}
        \item Were there any targets or potential targets of investigation that did not occur due to perceived lack of legal viability? 
    \end{enumerate}
    \item To what extent do you have the capacity to conduct fairness testing?
    \begin{enumerate}
        \item Are there any practical barriers to leveraging that capacity?
        \item If there is not enough capacity, what happens?
    \end{enumerate}
    \item What other “tools” are available to your agency besides enforcement or supervision? Such as rulemaking, publishing policy statements, circulars, etc.
\end{enumerate}

\noindent \textbf{Q5} What sort of documentation would you expect a firm to create and maintain about their efforts and decisions? What do you expect companies to document and justify the business necessity of their algorithmic systems?
\begin{enumerate}
    \item If yes, could you describe what you’d ideally expect to see in a search?
    \item If none, could you describe what you’d expect to see in a search? 
\end{enumerate}

\vspace{2mm}
\noindent \textbf{Q6} Are there any circumstances that you recall where there was a disagreement between a regulator and a covered entity about whether a viable LDA exists a viable LDA exists? What was the source of the disagreement and how did it resolve?

\noindent \textit{For the next few questions, we’re focused on what the agency has been doing to get companies to actually meet those expectations.}

\noindent \textbf{Q1} In the first part of our conversation, we talked about things you saw private firms do. In the second part of our conversation, we talked about what the regulatory expectations were. In your opinion, what is the agency doing/what did the agency do to ensure that companies meet those expectations? 

\noindent \textbf{Q2} In particular, In your experience, what do you think has hindered regulatory efforts to clarify that companies should be regularly testing their models for DI and searching for LDAs?
\begin{enumerate}
    \item Specifically, do you think the interplay between disparate treatment and impact has done so?
    \item To what extent have organizations in your jurisdictions noted or sought guidance around challenges related to the use of demographic data for measurement or mitigation of discrimination?
\end{enumerate}

\noindent \textbf{Q3} Do you believe that further clarification/standardization of requirements would motivate covered entities to be more proactive in searching for LDAs?

\noindent \textbf{Q4} In your experience, is there certain guidance or other ``papers'' that companies have asked you to provide to clarify expectations? 
\begin{enumerate}
    \item When writing guidance, what factors go into the specificity with which these guidances are written, especially with regards to more technical concepts like bias testing and LDAs?
\end{enumerate}

\noindent \textbf{Q5} To what extent, in your experience, do firms react to ``sub-regulatory'' documents, e.g., blogposts or public statements? <Priming examples below>
\begin{enumerate}
    \item Performance cost: accuracy changes
    \item Startup cost: having necessary infrastructure/ people in place
    \item Search cost: training runs
    \item Deployment cost: changing the deployment to the new model, and the cost of running that model; retraining people to use the new model
    \item Human cost: cost of making the decision to change the model, and cost of ``people-hours'' (time spent, salary) on conducting the search
\end{enumerate}

\noindent \textbf{Closing thoughts}

\noindent \textbf{Q1} How do you or your agency keep up to date on the state of the art regarding algorithmic testing, LDAs, and civil rights law?

\vspace{2mm}
\noindent \textbf{Wrap up}

\noindent \textbf{Q1} Is there anything you wanted to say that you didn’t get the chance to?

\noindent \textbf{Q2} Is there anyone else you could connect us with that might be a good fit for our interview?

\section{Codebook}
\label{app:codebook}
Our codebook can be found in Table~\ref{fig:codebook}.

\begin{table*}[]
\small
\begin{tabularx}{\linewidth}{|X|X|}
\hline
\textbf{Code/}Sub-code & \textbf{Code/}Sub-code\\ \hline
\textbf{Policies (rules)} & \textbf{Documentation} \\ \hline
{\begin{tabularx}{\linewidth}[t]{@{}X@{}}
Trigger for assessment\\ 
Assessment policy\\
Trigger for mitigation\\ 
Mitigation policy\\ 
Deployment policy\end{tabularx}} 
& 
 {\begin{tabularx}{\linewidth}[t]{@{}X@{}}
Documentation of policy\\ 
Assessment report\\ 
Mitigation report\\ 
Audience and access\\ 
Attorney client privilege\end{tabularx}} \\ \hline
\textbf{Procedures} & \textbf{Regulation from Regulators' Point of View}  \\ \hline
{\begin{tabularx}{\linewidth}[t]{@{}X@{}}
Access to and availability of data\\ 
Model development\\ 
Model validation\\ 
Demographic data\\ 
Bias testing\\ 
Bias mitigation\\ 
Trade-offs\end{tabularx}} &

{\begin{tabularx}{\linewidth}[t]{@{}X@{}}
Expectations\\ 
Regulatory clarity\\ 
Promote compliance\\ 
Reasonableness assessment\\ 
Legal basis\\ 
Enforcement \\
Willingness to tolerate experimentation\end{tabularx}}\\ \hline

\textbf{Processes (Practice/standards)} & \textbf{Organizational Dynamics}\\ \hline

{\begin{tabularx}{\linewidth}[t]{@{}X@{}}
Feature review process\\ 
Disparity review process\\ 
Business necessity\\ 
Mitigation strategy\\
Determining what counts as a viable alternative\\
Model deployment\end{tabularx}} & 
   {\begin{tabularx}{\linewidth}[t]{@{}X@{}}
Decision-making authority\\ 
Risk tolerance\\ 
Division of labor\\ 
Team structure and expertise\\ 
Fostering buy-in\\ 
Market position\\ 
Cross-functional interactions/collaborations/conflicts\end{tabularx}}\\ \hline
\textbf{Regulation from Entities' Point of View}                                             & \textbf{Financial Cost} \\ \hline
{\begin{tabularx}{\linewidth}[t]{@{}X@{}}
What laws are thought to apply\\ 
Perceived expectations of regulators, both now and in the future\\ 
Regulatory clarity\\ 
Legal risk\\ 
Political landscape\end{tabularx}}
& 
\\ \hline
\end{tabularx}
\caption{The codebook resulting from our reflexive thematic analysis, organized by code in bold, followed by its sub-codes.}
\label{fig:codebook}
\end{table*}

\end{document}